\documentclass{epl}

\newcommand{\avk}{\langle k \rangle}

\title{Synchronization of Kuramoto Oscillators in Scale-Free Networks} 

\shorttitle{Synchronization of Kuramoto Oscillators in SF Networks} 

\author{Yamir Moreno\inst{1,2}\thanks{E-mail:\email{yamir@unizar.es}}
\and Amalio F. Pacheco\inst{1,2}}

\shortauthor{Y. Moreno \etal}

\institute{
\inst{1} Departamento de F\'{\i}sica Te\'orica, Universidad de
Zaragoza, Zaragoza 50009, Spain\\
\inst{2} Instituto de Biocomputaci\'on y F\'{\i}sica de Sistemas
Complejos (BIFI), Universidad de Zaragoza, Zaragoza 50009, Spain
}


\begin{document}

\maketitle

\begin{abstract} 
In this work, we study the synchronization of coupled phase
oscillators on the underlying topology of scale-free networks. In
particular, we assume that each network's component is an oscillator
and that each interacts with the others following the Kuramoto
model. We then study the onset of global phase synchronization and
fully characterize the system's dynamics. We also found that the
resynchronization time of a perturbed node decays as a power law of
its connectivity, providing a simple analytical explanation to this
interesting behavior.
\end{abstract}

The behavior of an isolated generic dynamical system in the long-term
limit could be described by stable fixed points, limit cycles or
chaotic attractors \cite{sbook}. We have also learned in recent years
that when many of such dynamical systems are coupled together, new
collective phenomena emerge. In this way, the study of regular
networks of dynamical systems have shed light on a number of natural
phenomena ranging from earthquakes to ecosystems and living organisms
\cite{turcotte,levin,winfree}. One of the most fascinating phenomena
in the behavior of complex dynamical systems made up of many elements
is the spontaneous emergence of order and the phenomenon of collective
synchronization \cite{pik}, where a large number of the system's
constituents forms a common dynamical pattern, despite the intrinsic
differences in their individual dynamics. Of recent interest are a
plenty of biological examples that have become accessible at least
numerically with the advent of modern computers \cite{strogatz}.

On the other hand, it has been recently shown that many biological
\cite{ref5,ref6}, social \cite{pnas}, and technological \cite{alexei}
systems exhibit an intricate pattern of interconnections in the form
of complex networks \cite{strogatz}. This structural complexity cannot
be described by the couplings of a regular network. In order to
characterize topologically these complex networks, one computes the
probability, $P(k)$, that any given element of the network has $k$
connections to other nodes. Interestingly, many real-world networks
such as the Internet, protein interaction networks and social webs
\cite{book1} can be well approximated by a power-law connectivity
distribution, $P(k)\sim k^{-\gamma}$. Additionally, they are
characterized by the existence of key nodes which drastically reduce
the average distance between nodes, the so-called small-world
property \cite{ws98}.

Most of the studies performed so far have scrutinized the structure of
complex networks and studied prototype models ran on top of these
networks \cite{siam}. Their peculiar topological properties have been
shown to lead to radical changes when dynamical processes such as
epidemic spreading and percolation phenomena are studied on top of
complex heterogeneous networks
\cite{newman00,pv01a,moreno02,av03}. However, for biological and other
applications, it would be relevant to consider the nodes of a given
network as dynamical systems with own identity. Examples of such
dynamical systems are ensembles of coupled and pulse-couple
oscillators with and without time delay, widely used because of their
relevance to natural systems such as chirping crickets and flashing
fireflies, among others \cite{nr1,nr2}. Besides, there are several
studies where the conditions for complete synchronization in complex
networks are scrutinized \cite{pecora,wang,niki}.

In this paper, we numerically study the synchronization of coupled
phase oscillators following Kuramoto's model \cite{k1,k2} on the
underlying topology of an scale-free (SF) network. We report on the
system dynamics by computing the conventional order parameter. The
onset of synchronization is found at a nonzero value of the coupling
strength. Remarkably, the transition from a desynchronized to a
synchronous state can be characterized with the mean-field exponent
found for globally-coupled oscillators. The heterogeneous character of
the network allows us to explore the robustness of the synchronized
state under a single perturbation as a function of the connectivity of
the perturbed oscillator. Interestingly, we found that the more
connected a node is, the more stable it is.

Let us consider an SF network where each node $i$ $(i=1,\cdots,N)$ is a
planar rotor characterized by an angular phase, $\theta_i$, and a
natural or intrinsic frequency $\omega_i$. Two oscillators interact if
they are linked together by an edge of the underlying network. The
individual dynamics of the $i$th node is described by
\begin{equation}
\frac{d\theta_i}{dt}=\omega_i+\lambda\sum_{j \in I(i)}\sin(\theta_j-\theta_i)
\label{eq1}
\end{equation}
where $I(i)$ is the set of neighbors of the rotor $i$ as dictated by
the architecture of the network and $\lambda$ is the coupling
strength, identical for all edges. The natural frequencies and the
initial values of $\theta_i$ are randomly drawn from a uniform
distribution in the interval $(-1/2,1/2)$ and $(-\pi,\pi)$,
respectively. On the other hand, in order to produce SF networks we
have used the BA procedure \cite{bar99}. In this model, starting from
a set of $m_0$ nodes, one preferentially attaches at each time step a
newly introduced node to $m$ older ones ($m=3$ has been set). The
procedure is repeated many times and a network with a power law degree
distribution $P(k)\sim k^{-\gamma}$ with $\gamma=3$ and average
connectivity $\langle k \rangle=2m=6$ builds up. This network is a
clear example of a highly heterogeneous network because the degree
distribution has unbounded fluctuations when $N\rightarrow\infty$.

The original Kuramoto model corresponds to the simplest case of
globally coupled (all-to-all), equally weighted oscillators where the
coupling strength $\lambda=K/N$ to ensure that the model is well
behaved in the thermodynamic limit \cite{k1,k2}. The onset of
synchronization occurs at a critical value of the coupling strength,
$K_c=2/\pi g(\omega_0)$, where $g(\omega)$ is the distribution from
which the natural frequencies are drawn evaluated at the mean
frequency $\omega_0$. The second-order phase transition is
characterized by the following order parameter
\begin{equation}
r(t)=\left|\frac{1}{N}\sum_{j=1}^{N}e^{i\theta_j(t)}\right|
\end{equation}
which behaves when both $N\rightarrow\infty$ and $t\rightarrow\infty$
as $r\sim (K-K_c)^{\beta}$ for $K\ge K_c$ being $\beta=1/2$.

\begin{figure}[t]
\begin{center}
\onefigure[width=3.2in,angle=-90]{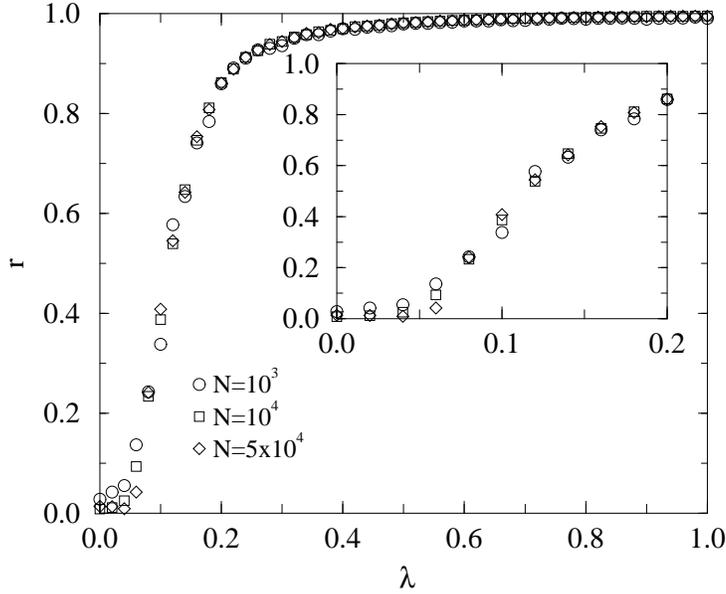}
\end{center}
\caption{Coherence $r$ as a function of $\lambda$ for several system
  sizes. The onset of synchronization occurs at a critical value
  $\lambda_c=0.05(1)$. Each value of $r$ is the result of at least 10
  network realizations and 1000 iterations for each $N$. The inset is
  a zoom around $\lambda_c$.}
\label{figure1}
\end{figure}

In order to inspect the dynamics of the $N$ oscillators on top of
complex topologies, we have performed extensive numerical simulations
of the model in BA networks. Starting from $\lambda=0$, we increase at
small intervals its value. Then, we integrate the equations of motion
Eq.\ (\ref{eq1}) over a sufficiently large period of time (at least
$10^4$ integration steps) to ensure that the system is in a stationary
state, and the order parameter $r$ is computed. The procedure is
repeated gradually increasing $\lambda$ until the system evolves to a
state of collective phase synchronization.

In the case of random SF networks the global dynamics of the system is
qualitatively the same as for the original Kuramoto model as shown in
Fig.\ \ref{figure1} for several system sizes. As the coupling is
increased from small values, the strength of the interactions is not
enough to break the incoherence produced by their individual
dynamics. This behavior persists until a certain critical value
$\lambda_c$ is crossed. At this point some elements lock their
relative phase and a cluster of synchronized nodes comes up. This
constitutes the onset of synchronization. Beyond this value, the
population of oscillators is split into a partially synchronized state
made up of oscillators locked in phase that adds to $r$ and a group of
nodes whose natural frequencies are too spread as to be part of the
coherent pack. Finally, after further increasing the value of
$\lambda$, more and more nodes get entrained around the mean phase and
the system settles in a completely synchronized state where
$r\approx1$.

\begin{figure}[t]
\begin{center}
\onefigure[width=3.2in,angle=-90]{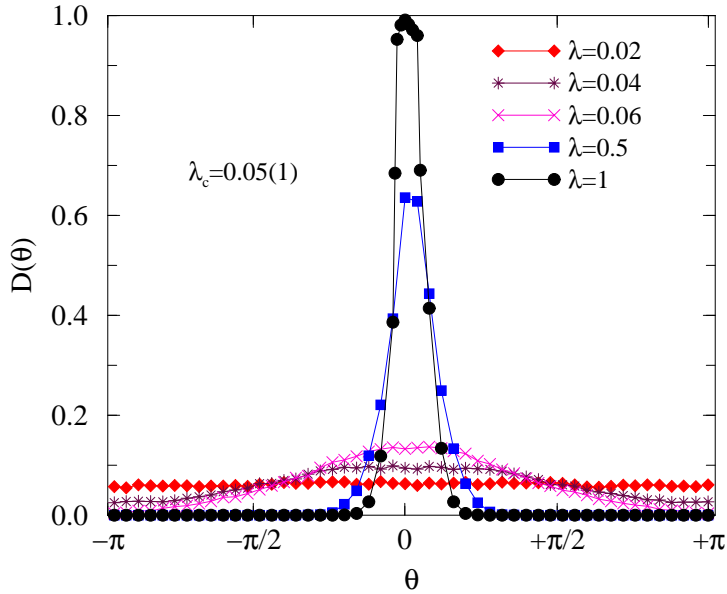}
\end{center}
\caption{Normalized phase distributions $D(\theta)$ for different
  values of the control parameter $\lambda$. The curves depicted
  correspond to values of $\lambda$ below, near and above $\lambda_c$
  as indicated. The network is made up of $N=10^4$ nodes.}
\label{figure2}
\end{figure}

This picture is clearly illustrated in Fig.\ \ref{figure2}, where we
have plotted the distributions of phases $D(\theta)$ for five
different values of $\lambda$ between $0$ and $1$. Below the critical
point the phases are uniformly scattered through the entire interval
$(-\pi,\pi)$. When $\lambda > \lambda_c$ the distribution shrinks
around the mean value $\theta=0$ and the dispersion gets smaller as
$\lambda$ grows, signaling that the system is in a synchronous state.

\begin{figure}[t]
\begin{center}
\onefigure[width=3.2in,angle=-90]{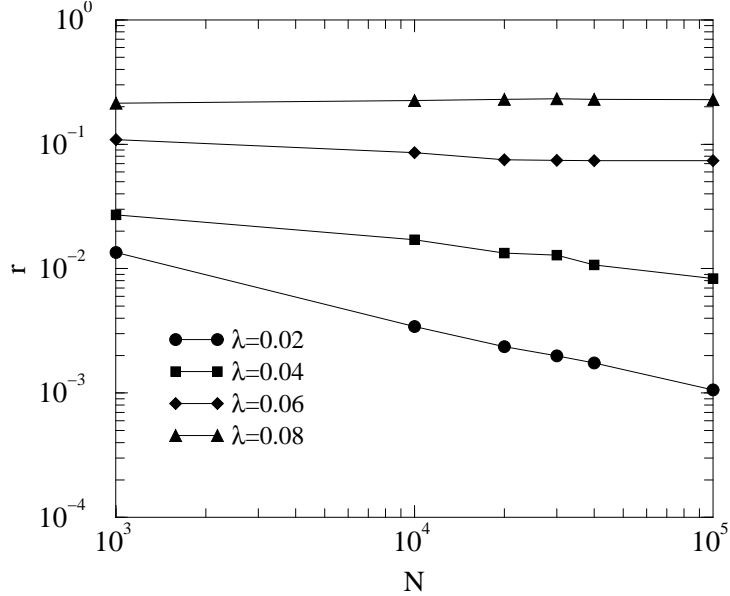}
\end{center}
\caption{Finite-size scaling analysis used to determine the value of
the critical point. The curves suggest the existence of a critical
point above $0.04$. The values of $r$ are collected after $10^4$
integration steps and averaged over a time window of $100$ additional
integration steps. Finally, each point is averaged over at least 10
network realizations and $10^3$ different initial conditions.}
\label{figure3}
\end{figure}

Now we proceed to investigate the critical parameters of the system
dynamics. First, we should determine with precision the exact value of
$\lambda_c$. This is not an easy task because there are several
sources of heterogeneity and averages should be properly taken $-$from
network realizations to the initial distributions of $\theta_i$ and
$\omega_i-$ through lengthly numerical calculations. Additionally,
finite-size effects come into play. We have determined $\lambda_c$ and
studied the critical behavior near the synchronization transition by
means of a standard finite-size scaling (FSS) analysis
\cite{marro}. For a given network size $N$, we have that no
synchronization is attained below $\lambda_c$ and that $r(t)$ decays
to a small residual value of size $O(1/\sqrt{N})$. Hence the critical
point may be found by examining the $N$-dependence of
$r(\lambda,N)$. For $\lambda < \lambda_c$ (sub-critical regime), the
stationary value of $r$ falls off as $N^{-1/2}$, while for $\lambda >
\lambda_c$, $r$ approaches a nonzero, stationary value as
$N\rightarrow\infty$ though still with $O(1/\sqrt{N})$
fluctuations. In this way, plots of $r$ versus $N$ as that of Fig.\
\ref{figure3}, allow us to locate the critical point $\lambda_c$.

\begin{figure}[t]
\begin{center}
\onefigure[width=3.2in,angle=-90]{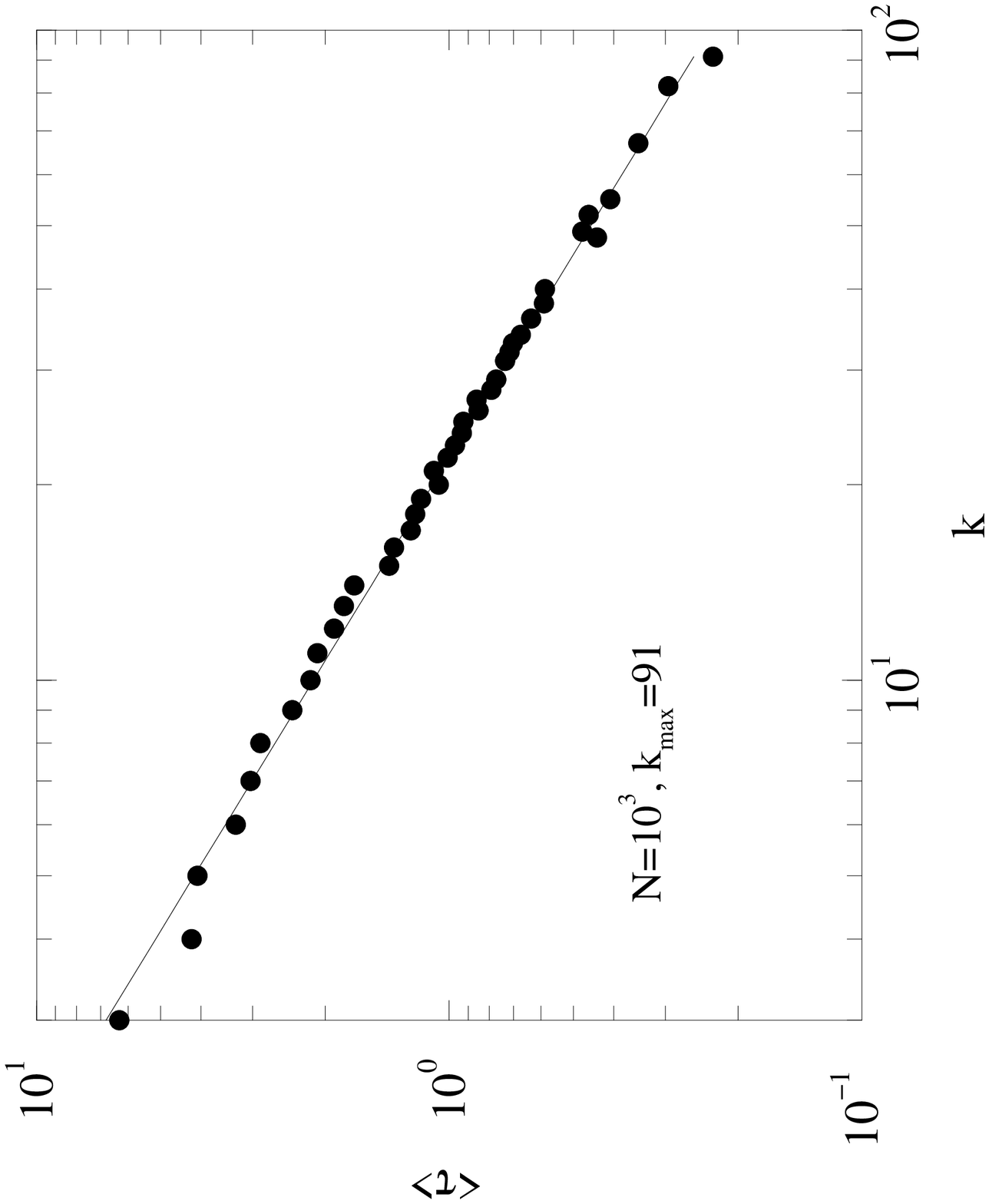}
\end{center}
\caption{Log-Log plot of the dimensionless average time $\langle \tau
\rangle$ it takes for a node of connectivity $k$ to be back in the
synchronous state after being perturbed. The least square fit to the
data gives for the exponent $\nu=0.96(1)$. The results were averaged
over 10 network realizations and 500 perturbations for each
$k$. $\lambda$ is set to $0.4$. See the text for details on the
definition of $\tau$.}
\label{figure4}
\end{figure}

  Following the FSS procedure, our best estimate gave a value for the
critical coupling strength $\lambda_c=0.05(1)$. Besides, we found that
$r\sim(\lambda-\lambda_c)^{\beta}$ above the critical point with
$\beta=0.46(2)$ indicating that the square-root behavior typical of
the mean-field version of the model (all-to-all architecture) also
holds for SF networks. In addition, it is worth stressing the very
existence of a critical point. This is the opposite to what has been
found when percolation or epidemic spreading models are ran on top of
complex heterogeneous networks even if correlations are taken into
account \cite{newman00,pv01a,moreno02,av03,moreno03}. Furthermore, the
critical point shifts to the left when the average connectivity $\avk$
of the underlying network increases, but it is always distinctly
different from zero. Other numerical calculations (not shown) indicate
that the relation $\lambda_c^{\avk}\cdot\avk$ is roughly constant when
$\avk$ varies, which implies that there is a non-trivial critical
point even in the infinite size limit \cite{mvp2}.

Once we have characterized the emergence of spontaneous
synchronization, we look at the stability and robustness of the
synchronous state. The most interesting and influential topological
property of complex heterogeneous networks is that the fluctuations of
the connectivity distribution are unbounded. In other words, there is
a clear distinction between the nodes according to their
connectivities. From a practical point of view $-$and worth taking
into account as a design principle in natural or artificial
networks$-$, it would be particularly relevant that the most highly
connected nodes were also the most robust with respect to
synchronization when they were perturbed. We have computed the average
time $\langle\tau\rangle$ it takes for a node to be again in the
synchronized cluster as a function of its connectivity $k$ after being
perturbed and put out of synchronization. This is done by assigning to
a randomly choosen node $i$ of connectivity $k_i$ a new phase
$\theta_i$ which differs in $-\pi$ to its synchronization value. In
this way, all nodes are perturbed in the same amount and one may
calculate the time it takes for $i$ to recover from the
perturbation. Note that as $\lambda$ is high enough, the oscillator
$i$ ends up in the synchronous state with the same $\theta_i$ it had
before being perturbed. The results are drawn in Fig.\ \ref{figure4},
where a clear power-law dependency $\langle\tau\rangle\sim k^{-\nu}$,
with $\nu=0.96(1)$, can be identified. Hence, the more connected a
node is, the more stable it is. The power-law behavior points to an
interesting result, namely, it is more easy for an element with high
$k$ to get locked in phase with its neighbors than for a node linked
to just a few others.  Furthermore, the destabilization of a hub does
not destroy the synchrony of the group it belongs to. Instead, it
works the other way around, the group formed by the hub's neighbors
recruits it again.

This behavior and the dependency $\langle\tau\rangle\sim k^{-1}$ may
be understood by the following simple argument. As we are perturbing a
single node $i$ and this perturbation, $\xi_i$, is small, we can
assume that it affects only the first neighbors of the perturbed
node. Hence, the stability analysis can be locally reduced to the
problem of how such a perturbation relaxes in a star topology (a
single perturbed hub attached to $k\gg1$ oscillators). This
approximation is particularly suitable in random networks with
arbitrary degree distributions (like the BA one) because the
probability of finding loops (tringles, cycles, etc) is small and
vanishes as $N$ grows. Linearization of Eq.\ (\ref{eq1}) for this
configuration leads to $\xi_i^{\eta}=\xi_i^{\eta}(0)e^{\eta_it}$,
where $\eta_i$ is the eigenvalue corresponding to the oscillator
$i$. Henceforth, the times $\langle\tau\rangle$'s are given by the
inverse of the eigenvalues, which for a star configuration are
$\eta_i=-1$, for $i=1, \cdots, N-2$ and
$\eta_{N-1}=\eta_{hub}=-N=-k_{hub}-1$ \cite{pecora2}. In other words,
the fastest relaxation rate in a start topology corresponds to the hub
and goes like $1/k_{hub}$ for $k_{hub}\gg 1$, while the rest of the
oscillators all have the same relaxation times. Additionally, the
eigenvector associated to the eigenvalue of the hub indicates that it
moves a lot while the other nodes change very little. Finally, if we
superpose the effects of many perturbations to different nodes (each
one being a $k_{hub}+1$ star), it comes out that each of them
contributes with $1/k_{hub}$, $k_{hub}=k_{min}, \cdots, k_{max}$ to
the $\langle\tau\rangle$ dependency with $k$, leading to the law
$\langle\tau\rangle\sim k^{-1}$ depicted in Fig.\ \ref{figure4}.

In summary, we have studied the synchronization of Kuramoto
oscillators on top of complex scale-free networks. We have found that
the onset of synchronization occurs at a nonzero value, though small,
with a critical exponent around $0.5$. We also found that when the
synchronous state has been attained, highly connected nodes are more
robust under perturbations and they are recovered in a time which
depends on their degree as a power law with exponent close to
$-1$. This law support the suggestion that the actual topology of
scale-free networks may be a result of some kind of optimization
mechanism at a local scale, optimizing in this case the fitness for
synchronization of highly connected nodes. This question as well as
the influence of time delay and noise on the synchronization of
complex networks make it necessary the study in more details the
connection between graph theory, dynamics on networks and nonlinearity
in future modeling of complex networks.

\acknowledgments 

We thank J.\ G\'omez-Garde\~nes for discussions and two anonymous
referees for comments and suggestions that greatly improved the
manuscript. Y.\ M.\ is supported by MCyT through the Ram\'on y Cajal
program. This work has been partially supported by the Spanish DGICYT
project BFM2002-01798.

\end{document}